\def\year{2021}\relax
\documentclass[letterpaper]{article} % DO NOT CHANGE THIS
\usepackage{aaai21}  % DO NOT CHANGE THIS
\usepackage{times}  % DO NOT CHANGE THIS
\usepackage{helvet} % DO NOT CHANGE THIS
\usepackage{courier}  % DO NOT CHANGE THIS
\usepackage[hyphens]{url}  % DO NOT CHANGE THIS
\usepackage{graphicx} % DO NOT CHANGE THIS
\usepackage{natbib}  % DO NOT CHANGE THIS AND DO NOT ADD ANY OPTIONS TO IT
\usepackage{caption} % DO NOT CHANGE THIS AND DO NOT ADD ANY OPTIONS TO IT
\usepackage[scaled=.95]{inconsolata}
\usepackage{enumitem}
\setlist{noitemsep}
\usepackage[hang,flushmargin]{footmisc}
\usepackage{array}

\title{Media Cloud: Massive Open Source Collection of Global News on the Open Web}
\author {
    Hal Roberts,\textsuperscript{\rm 1,2}
    Rahul Bhargava,\textsuperscript{\rm 3,2}
    Linas Valiukas,\textsuperscript{\rm 2}
    Dennis Jen,\textsuperscript{\rm 2}
    Momin M. Malik,\textsuperscript{\rm 1}\\
    Cindy Sherman Bishop,\textsuperscript{\rm 2}
    Emily B. Ndulue,\textsuperscript{\rm 2}
    Aashka Dave,\textsuperscript{\rm 4} 
    Justin Clark,\textsuperscript{\rm 1} 
    Bruce Etling,\textsuperscript{\rm 1,2} \\
    Robert Faris,\textsuperscript{\rm 5,1,2} 
    Anushka Shah,\textsuperscript{\rm 6,7}
    Jasmin Rubinovitz,\textsuperscript{\rm 7}\thanks{Now at Google.}
    Alexis Hope,\textsuperscript{\rm 7} 
    Catherine D'Ignazio,\textsuperscript{\rm 8}\\
    Fernando Bermejo,\textsuperscript{\rm 2,1}
    Yochai Benkler,\textsuperscript{\rm 9,1,2}
    Ethan Zuckerman\textsuperscript{\rm 10,1,2}
\\
}
\affiliations {
    \textsuperscript{\rm 1} Berkman Klein Center for Internet \& Society, Harvard University\\
    \textsuperscript{\rm 2} Media Cloud\\
    \textsuperscript{\rm 3} College of Arts, Media and Design, Northeastern University\\
    \textsuperscript{\rm 4} School of Information \& Library Science, University of North Carolina at Chapel Hill\\
    \textsuperscript{\rm 5} Shorenstein Center on Media, Politics and Public Policy, Harvard University\\
    \textsuperscript{\rm 6} Civic Studios\\
    \textsuperscript{\rm 7} Media Lab, Massachusetts Institute of Technology\\
    \textsuperscript{\rm 8} Department of Urban Studies and Planning, Massachusetts Institute of Technology\\
    \textsuperscript{\rm 9} Harvard Law School, Harvard University\\
    \textsuperscript{\rm 10} School of Public Policy, University of Massachusetts Amherst\\
    hroberts@cyber.harvard.edu, 
    r.bhargava@northeastern.edu, 
    \{linas, dennis\}@mediacloud.org, 
    momalik@cyber.harvard.edu, \\
    \{cindy, ebndulue\}@mediacloud.org, 
    aashkad@email.unc.edu, 
    \{jclark, betling, rfaris\}@cyber.harvard.edu,
    \{anushka18, \\jasrub\}@gmail.com, 
    \{ahope, dignazio\}@mit.edu, 
    fernando@mediacloud.org, 
    ybenkler@law.harvard.edu, 
    ethanz@umass.edu
}

\usepackage[pages=some]{background}

\backgroundsetup{placement=top,
contents={Appearing in Proceedings of the Fifteenth International AAAI Conference on Web and Social Media (ICWSM 2021)},
color=black,
angle=0,
opacity=1,
scale=.8,
vshift=-47,
firstpage=true
}

\begin{document}

\maketitle

\begin{abstract}
We present the first full description of Media Cloud, an open source platform based on crawling hyperlink structure in operation for over 10 years, that for many uses will be the best way to collect data for studying the media ecosystem on the open web. We document the key choices behind what data Media Cloud collects and stores, how it processes and organizes these data, and its open API access as well as user-facing tools. We also highlight the strengths and limitations of the Media Cloud collection strategy compared to relevant alternatives. We give an overview two sample datasets generated using Media Cloud and discuss how researchers can use the platform to create their own datasets.
\end{abstract}

\section{Introduction}
Researchers have long studied the production and dissemination of news to understand its impact on democracy, beliefs and behaviours. These efforts have included large scale projects to manually review news content \cite{kayser1953}, theories of how ``themes'' of news move subjects into and out-of public favor \cite{fishman1978}, sociological accounts of newsroom decision making \cite{gans1979}, analysis of news agendas to help identify areas ripe for public health messaging \cite{dorfman2003}, and more. While the digitization of news has broadly made many sorts of research newly possible, the volume of content and newer trends toward proprietary platforms have significantly increased the barriers for researchers hoping to build on the scholarly traditions of studying attention, representation, influence, and language in online news. 

In addition, the democratization of content authoring has significantly expanded the number of different news sources reporting about any topic or geographic area. The process of discovering which online media sources to study has become more involved; how does one identify the set of media outlets that exist and decide which are influential? Furthermore, once a set of media sources has been created, collecting their content introduces many technological difficulties. 

Many researchers rely on commercial data stores to solve these problems of media source discovery and content collection, but those stores are often limited by cost, timeliness, geographic focus, linguistic diversity, and content restrictions. Analysis falling outside of those stores requires significant technical and substantive effort. 

To address these concerns we created the Media Cloud project, an open source data corpus and suite of web-based analytic tools that support research into open web media ecosystems across the globe. This paper documents the architecture and datasets underlying Media Cloud. We provide some history of the project origins, document the data collection and metadata extraction processes, introduce the datasets Media Cloud allows its users to create, review methods of accessing those datasets, and summarize research the datasets allow. Researchers can create datasets via the API or web-based research support tools to study attention, representation, influence, and language about subjects of their own choosing. They may also download and run the Media Cloud code to create their own instances of the platform.

\section{About Media Cloud}
Since 2008, Media Cloud has collected a total of over 1.7 billion stories. As described below, these data are available to the public via an open source code base available on GitHub (\url{https://github.com/mediacloud}), a suite of free web tools (\url{https://tools.mediacloud.org}), and an extensive open API (\url{https://mediacloud.org/support}). Our project has a growing number of open-source contributors and collaborators outside of our core team. Media Cloud is limited by copyright restrictions from sharing the full text of news articles with external partners; instead, we surface metadata, data about text content of the documents, and URLs of documents indexed. After reviewing the history of Media Cloud, this section documents the data available in the underlying platform and how those data are generated. 

\subsection{History}
Work on Media Cloud began at Harvard's Berkman Klein Center for Internet \& Society in 2007 as a project to examine the influence of blogging on mainstream media coverage. Berkman researchers were using the blog search engine Technorati to identify popular topics in blogs and sought an up-to-date, open source alternative to LexisNexis to identify topical coverage in news media. 

Earlier software had been developed at Berkman to track a subset of topics in news media through keywords---the names of individual countries---to support the ``Global Attention Profiles'' (GAP) research project \cite{zuckerman2003}. While the GAP tool tracked hundreds of specific keywords by scraping the search engines of select news websites, Media Cloud was designed as a general solution, allowing researchers to track keywords across websites with RSS feeds. GAP maintained only a count of how many stories mentioned given keywords on a specific day, but Media Cloud provided a search engine, fully indexing the content of tracked pages. 

An internal debate within the Berkman Klein Center pushed the initial development of Media Cloud. Yochai Benkler's work on commons-based peer production suggested that permissionless online spaces for publication---blogs, online-native publications and wikis---could be more inclusive and diverse than traditional news media \cite{benkler2006}. Ethan Zuckerman argued that digital media was subject to strong homophily effects and would likely be less diverse than traditional news media \cite{zuckerman2013}. Media Cloud illuminated this debate with data: would `digitally native' media show greater topical and geographic diversity than traditional media? A 2012 Ford Foundation grant boosted the project, focusing on politically-motivated misinformation and the network architecture of agenda-setting in the public sphere \cite{benkler2013}.

Media Cloud became useful not only for the analysis of the architecture of large scale media ecosystems, but for analysis of the fine-grained flow of specific terms and ideas online over time. This affordance was facilitated by the system's ability to track story publication dates, as seen in work to understand the spread of news around Trayvon Martin's death in 2012 \cite{graeff2014}. 

Open and free access to its code, data, and tools has been a foundational principle for the project from the start. To make the platform more widely accessible, a team at the MIT Media Lab built interfaces to the Media Cloud datasets, usable by activists and journalists as well as researchers comfortable writing their own code to interact with the Media Cloud API. This expanded access led to calls to broaden the subject matter covered by Media Cloud, including the establishment of collections of news media in multiple countries outside the English-speaking world, beginning with collection of data about Brazil \cite{goncalves2014}. 

Since those beginnings, our global work has expanded significantly, producing a dataset of geographic collections of media sources that cover the large majority of countries in the world. Much of this diversification effort has been driven by funding partnerships with global grant-making non-profits. The Bill and Melinda Gates Foundation, the Ford Foundation, the MacArthur Foundation, the John H. and James L. Knight Foundation, the Robert Wood Johnson Foundation, the Open Society Foundations and others are heavily invested in assessing and influencing global media narratives on the topics they focus on. They provide significant funding to nonprofits specifically to impact local media narratives around issues such as contraceptive and family planning in Nigeria and sanitation in India. These foundations' support, and our own global research partnerships, have helped Media Cloud grow our international media coverage. The breadth of our collection differentiates the Media Cloud datasets from many commercial offerings. 

In addition to increasing the usability of the tools and the breadth of content coverage, the project has greatly expanded the kinds of data collected and the analysis performed on that data, in particular within the Topic Mapper tool. Led by a team at the Berkman Klein Center for Internet \& Society, the project grew from its initial scope of spidering and analyzing topics that were keyword-based and web-only, across a limited scope of about 10 thousand stories, to now managing web- and social media-seeded topics spanning the U.S. national media ecosystem and consisting of tens of millions of stories. 

\begin{figure*}[!ht]
    \centering
    \includegraphics[width=7in]{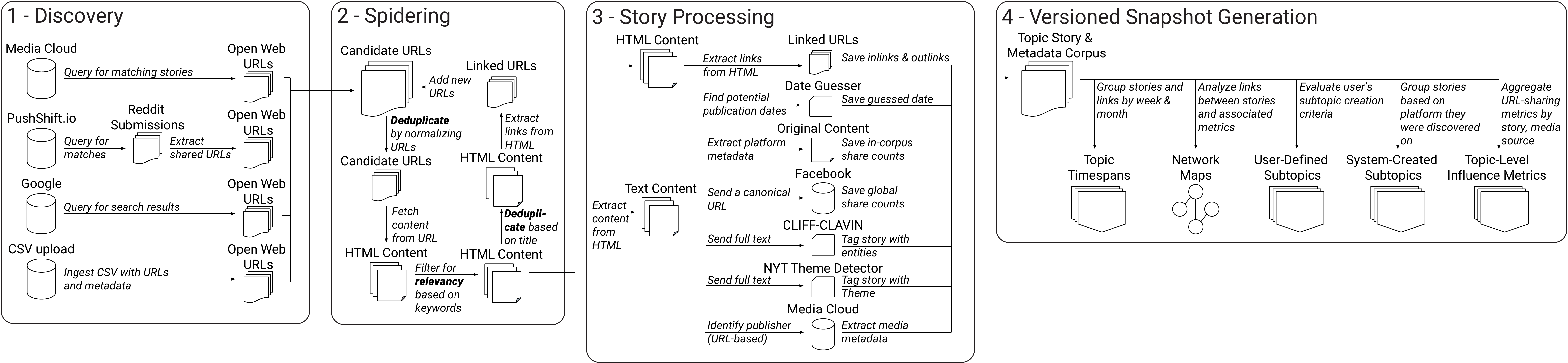}
    \caption{A schematic diagram of the Media Cloud topic-specific dataflow.}
    \label{fig:dataflow}
\end{figure*}

\subsection{Data Collection}
Media Cloud discovers and collects data through two systems: a syndicated feed crawler that ingests roughly a million stories per day from about 60 thousand media sources, and a topic-specific engine that recursively spiders hyperlinks to discover additional topic-specific content, capable of creating topic-specific datasets of more than 10 million stories. An overview of the system is given in figure (\ref{fig:dataflow}). 

\subsubsection{Regular Data Ingest}
The syndicated feed crawler ingests RSS, RDF, and Atom feeds associated with specific media sources. A media source is any publisher of online content, such as The New York Times or Gateway Pundit. Syndicated feeds for each media source are discovered through code that identifies feeds by: spidering two levels deep from the home page of the media source, finding any links that match a pattern indicating likely feeds (including terms like `RSS' or `XML'), downloading each of those URLs, and attempting to parse the resulting content as a syndicated feed to determine whether it has valid content.\footnote{\url{https://github.com/mitmedialab/feed_seeker}} All such validated feeds are added to the given media source. Also, a human ``collections curator'' regularly reviews collections and sources and manually adds missing feeds. 

Each feed is downloaded following a progressive backoff strategy. After a new story appears in a given feed, that feed is downloaded again within five minutes. If no new story is available, that feed is downloaded again in ten minutes, then twenty minutes, and so on, backing off on the frequency of update until the feed is only downloaded once a week. Once a new story is discovered in a given feed, its interval is reset to five minutes. 

The system parses the items from each downloaded feed and either: 1) matches each item to an existing story in the database by URL, GUID, or title, or 2) adds the unmatched item as a new story with the standard metadata provided by the feed (URL, GUID, title, publication date, description, and media source). To match by URL or GUID, the system uses a lossy URL normalization routine.\footnote{More details available at \url{https://mediacloud.org/news/2020/7/14/tech-brief-how-we-deduplicate-content}} To match by title, the system uses a normalization scheme based on the longest part of a title as separated by characters such as `\texttt{:}', `\texttt{|}', or `\texttt{-}'. 

The system downloads the HTML content for each story, extracts the substantive text using the publicly available \texttt{python-readability} package,\footnote{\url{https://github.com/buriy/python-readability}} parses the extracted text into individual sentences, removes duplicate sentences within the same media source and calendar day, and then stores the remaining concatenated sentences as the story text. The system then guesses the language of the story using the \texttt{cld2} library.\footnote{\url{https://github.com/CLD2Owners/cld2}} That story text is stored along with the language and other story metadata in a database to make it available for search through the platform API and various web tools. The system can also download and generate story text for podcasts (transcribed via the Google Speech-to-Text API). Finally, for each English story, the system also produces geotagging information using disambiguated CLIFF-CAVIN \cite{dignazio2014}, U.S. news subject topics based on New York Times topics modeled using Google News word2vec \cite{rubinovitz2017}, and entities generated by Stanford CoreNLP \cite{finkel2005}. 

As a principle, Media Cloud only downloads content that is freely available on the web. The vast majority of media sources make their content freely available to bot systems like Media Cloud, including even paywalled sites like those of the New York Times and the Washington Post. For the handful of sites that do not allow our bot access to full content, we index the content available. For example, the Wall Street Journal website provides the first paragraph of content, which is what we index. Likewise, many academic journals only provide open access to an abstract, which is what we index.

\begin{table*}[!ht]
\centering
\fontsize{9pt}{10pt} \selectfont
\begin{tabular}{l>{\raggedright\arraybackslash}p{2.6in}>{\raggedright\arraybackslash}p{2.9in}}
\hline
\textbf{Field} & \textbf{Search for} & \textbf{Example}\\ \hline
\texttt{{story\_id}} & a specific story by its unique Media Cloud id & \texttt{story\_id:12345}\\
\texttt{{media\_id}} & stories from a specific media source & \texttt{media\_id:12345}\\
\texttt{{publish\_date}} & stories published on a specific date and time & \texttt{publish\_date:2018-04-17 13:24:12}\\
\texttt{{publish\_day}} & stories published on a specific date & \texttt{publish\_day:2018-04-17 00:00:00}\\
\texttt{{publish\_week}} & stories published during a specific week\\
\texttt{{publish\_month}} & stories published during a specific month\\
\texttt{{publish\_year}} & stories published during a specific year\\
\texttt{{tags\_id\_stories}} & stories with a specific metadata tag & \texttt{tags\_id\_stories:9362290} (``Vladimir Putin'' tag)\\
\texttt{{tags\_id\_media}} & stories from a source with a specific metadata tag & \texttt{tags\_id\_media:32018496} (``digitally native source'' tag)\\
\texttt{{timespans\_id}} & stories from a specific timespan from a Topic & \texttt{timespans\_id:74322}\\ 
\texttt{{language}} & text in a language, by ISO-639-1 two-letter code & \texttt{language:en}\\ \hline
\end{tabular}
\caption{Fields included as indices in the Media Cloud database. See the online appendix in \url{https://arxiv.org/abs/2104.03702} for a description of data schema.}
\label{tab:fields}
\end{table*}

\subsubsection{Topic-specific Data Ingest and Spidering}
Media Cloud's Topic Mapper (\url{https://topics.mediacloud.org/}) operates independently from the feed collection system. A topic initially consists of a set of stories defined by a seed query, a set of media sources, and a date range. Topic Mapper collects stories for specific topics by searching the Media Cloud index for this initial batch of stories, and spiders out on the open web from hyperlinks in those stories to find other relevant stories. The resulting aggregate set of stories is added to the topic and acts as the seed set of URLs for the further spidering process. The topic spider uses the same normalized URL and title matching as the syndicated feed crawler to match spidered URLs to existing stories within the database and downloads unmatched URLs. This is shown in figure (\ref{fig:dataflow}). 

Whether a matching story is found in the database or the URL is novel and then downloaded, the content for each discovered story is matched against the seed query. The story is added to the topic only if it matches the seed query. For new stories, the system guesses a date using the story HTML,\footnote{\url{https://github.com/mediacloud/date_guesser}} a media source based on the URL hostname, and a title based on the HTML.  The system then processes the story through the same pipeline as feed-discovered stories. For each story added to the topic, the system parses hyperlinks from the extracted HTML of the story and adds them to the spidering queue. This spidering process is repeated for 15 rounds. 

The Topic Mapper can also seed topics with URLs through searches of social media platforms. For each supported platform, the system searches for all posts matching a given query, parses the hyperlinks out of the posts, and adds those URLs to the topic seed URLs. The system currently uses Brandwatch to search Twitter, Pushshift to search Reddit \cite{baumgartner2020} and verified Twitter accounts, and CSV uploads to support user-supplied platform searches (such as content downloaded from CrowdTangle). 

The Topic Mapper uses information from this discovery process to create influence metrics for stories and media. During the spidering process, all hyperlinks encountered between stories are made available in topic datasets. Those hyperlinks are also used to generate a media in-link count for each story and for each media source, where the media in-link count for a story is defined as the number of unique media sources linking to that story. Similarly for media sources, the media in-link count (for a source) is the number of unique media sources linking to any story within that media source. Those media in-link counts allow researchers to focus on the stories that have received the most attention in the link economy. The hyperlink data are also used to generate network visualizations of the media sources involved in a given topic, with the media sources as nodes and the hyperlinks as edges. The system also retrieves a Facebook share count for each story from the Facebook API and then sets the Facebook share count of each media source to be the sum of the source's story Facebook share counts. 

Any given topic can be analyzed as a subset of stories by specifying a query (as described below) that can restrict stories by text, language, or publication. For instance, an ``immigration" subtopic of an ``election topic" would include only stories mentioning immigration within the larger election topic.  Likewise, every topic can be analyzed within distinct weekly and monthly timespans, each of which consists of all stories published within a given date range or linked to by a story within the date range. And for topics that include a social media seed query, the system generates a subtopic that uses URL co-sharing by users as a replacement for hyperlinks for the network analysis (allowing, for instance, a network visualization of media sources that clusters together sources shared by the same users). The system also generates post, author, and channel counts for each of these social media seed queries. So a topic with a Reddit seed query will include a subtopic that allows analysis of co-sharing of URLs by Reddit users and also counts of the unique posts, channels, and authors that included each URL returned by the seed query. All of these modes of analysis are included within each topic's downloadable dataset. 

\subsection{Data Access}
Media Cloud runs on top of a scalable, containerized architecture currently consisting of around 80 docker containers running roughly 550 workers across a pool of servers. The core database is a PostgreSQL database that is responsible for storing nearly all of the system metadata. All story content is also indexed in a cluster of Apache Solr shards, which provides rapid text searching for the system. Below we summarize the multiple points of data access that this architecture allows for. 

\subsubsection{Query Language}
Queries are authored in a customized version of the basic Boolean query language supported by the Solr search engine. This allows for off-the-shelf inclusion of operators such as AND and OR, negation, proximity, word stemming (in multiple languages), word frequency thresholds, quoted strings, and more. Queries that are the length of long paragraphs are common for big research projects. Researchers can also filter by the fields included as indices in the database, shown in table (\ref{tab:fields}).

We have experimented with more advanced query tools such as query expansion and combining Boolean queries with trained models. These approaches are still deployed as one-offs, and are not yet available as general purpose tools within Media Cloud.

\subsubsection{API}
All of the data described above are available through an open API, which includes over 65 distinct end points. The public web tools, which are the primary research tools used by the Media Cloud team for its own research, are implemented entirely on top of this API, so all functionality in the front end tools is available via the API. The API is available both directly as a REST API with extensive documentation\footnote{\url{https://github.com/mediacloud/backend/blob/release/doc/api_2_0_spec}} and as a Python client we maintain and support.\footnote{\url{https://pypi.org/project/mediacloud/}} 

To support and ease adoption of API-based access for data science researchers, we have created a set of Jupyter Notebooks in the Python programming language. These notebooks provide examples of common operations such as listing top words used in stories matching a query, or accessing a topic dataset \cite{bhargava2020}.\footnote{\url{https://github.com/mediacloud/api-tutorial-notebooks}}

\subsubsection{Web-based Research Tools}
For those without programming experience, or those who simply want the convenience of a web-based research platform, we provide access to data through two analysis-focused web apps (Explorer and Topic Mapper), in addition to a website that functions as a library and management console for the media sources included in Media Cloud (Source Manager).\footnote{For screenshots, see online appendix in \url{https://arxiv.org/abs/2104.03702}.} For many researchers, these three apps are the only necessary access points to the Media Cloud dataset and its capabilities. 

Both Explorer and Topic Mapper enable researchers to investigate a sub-corpus of news content defined through a Boolean search including query text, media sources or collections, and a date range. Both apps present tools for analyzing attention (amount of coverage given to a keyword or topic), language (most frequently used terms), and representation (frequently mentioned people, organizations, and countries). Topic Mapper includes a deeper set of influence and linking metrics and research support features, as described previously. 

Explorer (\url{https://explorer.mediacloud.org/}) queries return in real time, allowing for rapid iteration on query construction. Researchers can use Explorer by itself to glean quick insights, and can use Topic Mapper (\url{https://topics.mediacloud.org/}) when asking deeper questions on influence: which media producers are doing most of the reporting on a topic, as well as the amount of referencing, citing, or sharing of the topic's media on social platforms by the general population. In addition to direct access to datasets via the API, the web tools provide links to CSV datasets underlying each visualization. 

Source Manager (\url{https://sources.mediacloud.org/}) provides researchers with information about individual news sources, and collections of news sources, within our system. Collection patterns, ingest rates, and other background information about the data contained are exposed to researchers. Critically, Source Manager tracks collection quality, so a researcher knows if data are incomplete for any range within the time period she is studying.

\section{Related Work}
There are three groups of platforms or tools that are similar in scope to Media Cloud. The first are news databases. Of these, the most comparable are LexisNexis, Google News, and mediastack. The second are event databases, of which the most comparable are GDELT and Event Registry. The third are crawler tools, of which the most comparable is IssueCrawler. We compare Media Cloud to each group in turn. 

LexisNexis is the most prominent commercial news database, providing archives of thousands of media sources which has made it the go-to tool for social science media research \cite{deacon2007}. It differs from Media Cloud in three important ways. The first is cost and accessibility. Media Cloud is free and open source, whereas LexisNexis is both expensive and proprietary. While some universities have subscriptions to LexisNexis, this may not include bulk API access without which a user must laboriously go through the web interface, selecting up to ten stories per page and downloading stories in batches of up to 100.

A second difference is coverage. LexisNexis is exhaustive for many sources, since it often has direct relationships with publishers. Particularly noteworthy is that it is a record of print archives in addition to online content, and it allows access to the full text content (images, graphics, audio, and video are not included) of media sources that may be behind paywalls or otherwise not on the open web. But it focuses almost exclusively on formal sources (professional media outlets); apart from a few blogs, informal sources are largely absent. Sites that do not publish in a news format, such as the World Health Organization and Wikipedia, are not covered at all. The third key difference is in story discovery. LexisNexis has no ability to go beyond the sources it covers, whereas Media Cloud uses spidering to iteratively follow hyperlinks, including to non-news sites such as Wikipedia. 

The lack of public documentation of LexisNexis also makes it difficult to freely compare to other tools. It claims coverage of 80 thousand sources, in 75 languages covering over 100 countries,\footnote{\url{https://internationalsales.lexisnexis.com/nexis-daas};\\archived snapshot at \url{https://perma.cc/XC5C-3XLS}} but further details as well as documentation of its API is not publicly available. Even if formal comparison is difficult without access, we emphasize the difference in use cases: for an authoritative search of formal news sources, LexisNexis is superior. Media Cloud is instead built for studying the media ecosystem on the open web, including hyperlink structure and non-formal sources. 

A more recent competitor is Google News (now just ``News API''). A 2008 study \cite{weaver2008} compared Google News and LexisNexis, finding inter-database agreement ranging from 29\% to 83\%, mostly due to LexisNexis excluding wire services. Today, News API boasts of access to 50 thousand sources.\footnote{\url{https://newsapi.org/};\\archived snapshot at \url{https://perma.cc/P63U-D8BJ}} It offers a free version, although with the significant restriction of only being able to search for content in the past three months. The commercial options are better but still limited: coverage there is in the past three years. News API focuses on providing streaming live, top headlines: for this it is superior to Media Cloud, although it is not an appropriate tool for research that seeks to consider older sources or collect more sources via snowball sampling. 

Lastly among news databases, mediastack (\url{https://mediastack.com/}) is a commercial product from the company apilayer, although it does allow for 500 free requests per month. Commercial options start from \$25/month for 10 thousand calls. mediastack covers ``general news, business news, entertainment news, health news, science news, sports news and technology news'' with both a live stream (updated every minute, and for which ``each of the sources used is monitored closely and around the clock for technical or content anomalies'') and a historical news data REST API, although without any description of how long this historical coverage extends or how it is collected (e.g. as an archive of its live stream, or something more). It does not have documentation of its collection process nor any academic papers we could find describing or making use of it, and only says that it covers over 7.5 thousand sources over more than 50 countries and in 13 languages. The countries are publicly listed (\url{https://mediastack.com/sources}), although the sources are not, which makes it difficult to compare to other news databases. 

Next, there are event databases \cite{wang2016}. First, GDELT, the Global Database of Events, Language and Tone \cite{leetaru2013}, also makes use of news stories on the open web (unlike other event databases, such as ICEWS), and is free and run from within academia. GDELT stores full-text data in some cases, but mainly focuses on collecting metadata. However, it is incompletely documented, and has received considerable criticism \cite{ward2013,weller2014,kwak2014a,kwak2014b,kwak2016,wang2016} around what its collection contains, with critics citing problems with duplication, lack of coverage of non-Western news sources, and dictionary-based pattern matching that makes it unable to make use of words not in its dictionary. It also does not include any spidering functionality for following hyperlinks beyond the sources it tracks. 

Event Registry (\url{https://eventregistry.org/}) was originally an academic project \cite{leban2014,rupnik2017} at the Institut Jo\v{z}ef Stefan (IJS) in Slovenia. However, in 2017 it was spun off into a commercial entity \cite{leban2020}, with the original tool no longer available. Recent academic papers published with Event Registry \cite[e.g.,][]{choloniewski2019,sittar2020} have been from IJS authors. It may have been an option for academic research until as late as 2019 \cite{colladon2020}, but at the time of writing the site did not list any academic or free access options. 

The paper initially describing the Event Registry \cite{leban2014} focused on \textit{events} as the end-goal, with news stories treated as a measurement instrument rather than the construct of interest. The tool described there sought to identify underlying events from a collection of articles using a clustering approach, and then extracted event metadata (location, date, parties involved) using custom named entity recognition and disambiguation. The paper describes relying on the IJS NewsFeed \cite{trampus2012}, also from the Institut Jo\v{z}ef Stefan. This seems to still be active (\url{http://newsfeed.ijs.si/}), although the page notes the stream is only internally accessible due to copyright issues. It also appears open source, although not updated since 2012.\footnote{\url{https://github.com/JozefStefanInstitute/newsfeed}}

In a 2016 comparison of GDELT and Event Registry \cite{kwak2016}, a rate limit on Event Registry made a direct comparison impossible. \citeauthor{kwak2016} did not attempt an article-by-article comparison, but found an overlap of about 13 thousand websites between GDELT's 63 thousand and Event Registry's 23 thousand, leading to a conclusion that the two offered different views of the world, without any clear way to tell how they might differ. 

The commercial tool currently described on the Event Registry website seems far more extensive than what was described in the \citeyear{leban2014} paper, and potentially than what was tested against GDELT in \citeyear{kwak2016}. Instead of relying on the IJS NewsFeed, Event Registry appears to have a custom collection (although perhaps built on the IJS NewsFeed) of ``over 150,000 diverse sources'' in over 40 languages that it has in fact spun out into a separate news API (\url{https://newsapi.ai/}). The site says that the tool allows clients to ``access news and blog content'' with search options by ``keyword, tags, categories, sentiment, location and more'' alongside the ability to ``group content into events'' that was the focus of the original paper. Articles are returned in groups of 100, with options to include the full body of text alongside metadata fields.\footnote{\url{https://eventregistry.org/documentation?tab=searchArticles}; archived snapshot at \url{https://perma.cc/TWN7-D2MR}} Beyond this, there is no information about what the tool is like or how much it costs; the website asks potential clients to get in contact for both demos and pricing options. 

Now that Event Registry is now a commercial tool that lacks public documentation or free access, direct comparison with Media Cloud is difficult. However, that means that for academic and non-commercial usage where access, transparency, and documentation are important, Media Cloud has a clear advantage. 

Lastly, there are web crawling tools. IssueCrawler \cite{rogers2006,rogers2010,rogers2019} is possibly the earliest \cite{rogers2000} and longest-running academic project to provide web crawling functionality for social science researchers. It has a strong theoretical motivation and backing based in science and technology studies \cite{rogers2000,marres2008,rogers2010,rogers2012}, has been used by researchers external to IssueCrawler for quite some time for studying conversations around civil society \cite{mcnally2005,siapera2007,chowdhury2013,schmidpetri2017}, and shares functionality of the spidering component of Media Cloud. However, on the technical side, its spidering is far more limited: IssueCrawler does not perform automatic deduplication, does not have its own internal seed set (previous sources of seed sets are now defunct), does not restrict stories by keyword, and requires repeated iterations of crawls to be done manually \cite{rogers2018}. It is free, although its source code does not appear to be available or open-source. It may be preferable for those who are seeking a minimal tool to do only spidering and want to do their own deduplication, filtering, and iteration in order to have fine-grained control, or who wish to use some of IssueCrawler's other link analysis tools.

Commercial tools such as Brandwatch (into which the academic-friendly tool Crimson Hexagon recently merged; Crimson Hexagon previously sold full Twitter data access to academics at a discounted rate), Zignal (which accesses social media data, and also gives subscribers access to LexisNexis), or Parse.ly offer some of the same functionality as Media Cloud. These commercial tools are focused on providing ready-made tools for analysis to non-academic clients, and lack documentation or access that would allow comparison with open-source and/or non-commercial tools. Since 2017, Event Registry also falls into this category. 

We also note that there exist one-off datasets, including those published in ICWSM \cite{horne2018}. These obviously differ from Media Cloud and other tools in that they are static.

\section{About the Data}

\subsection{Data Platform}
\begin{figure}[b]
    \centering
    \includegraphics[width=\columnwidth]{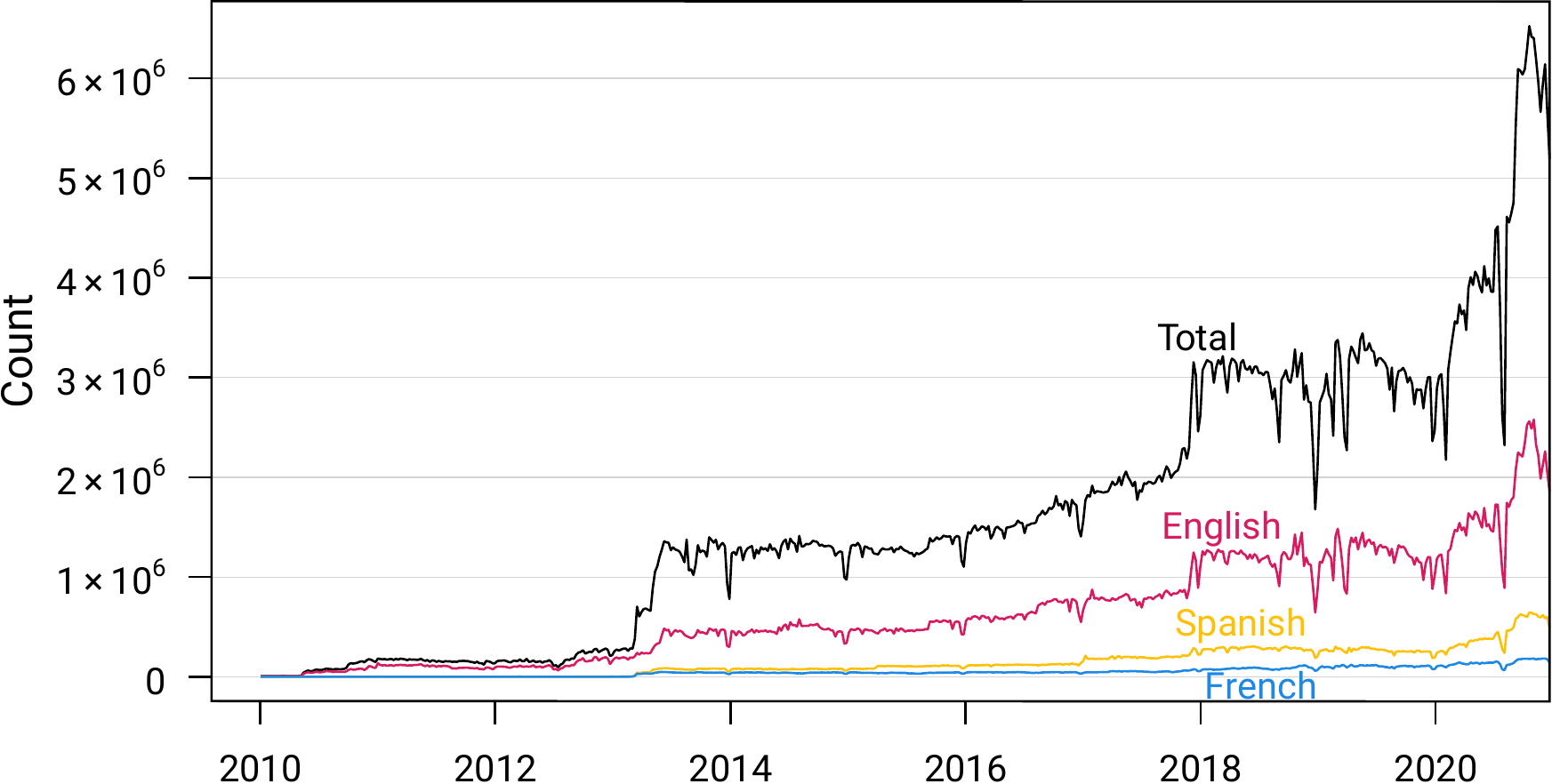}
    \caption{The total number of stories collected by Media Cloud per 7-day period (Sundays to Saturdays) since collection started in 2010, as well as the number of stories collected in English, Spanish, and French.}
    \label{fig:coverage}
\end{figure}

Media Cloud is built as an open platform for generating research datasets.  The underlying Media Cloud database includes over 1.7 billion stories belonging to 1.1 million distinct media sources, which the platform makes available through the tools and query language described above. The growth in the number of stories over time is shown in figure (\ref{fig:coverage}). While the entirety of the underlying data is merely a collection of media sources and topics in which the Media Cloud has been interested over the past decade, and so is not a useful representative sample of a general population of interest,  the API and web tools described above enable researches to create their own datasets based on their own interests. In this section, we describe two sample datasets generated using the platform's publicly available tools.

\begin{figure}[h]
    \centering
    \includegraphics[width=\columnwidth]{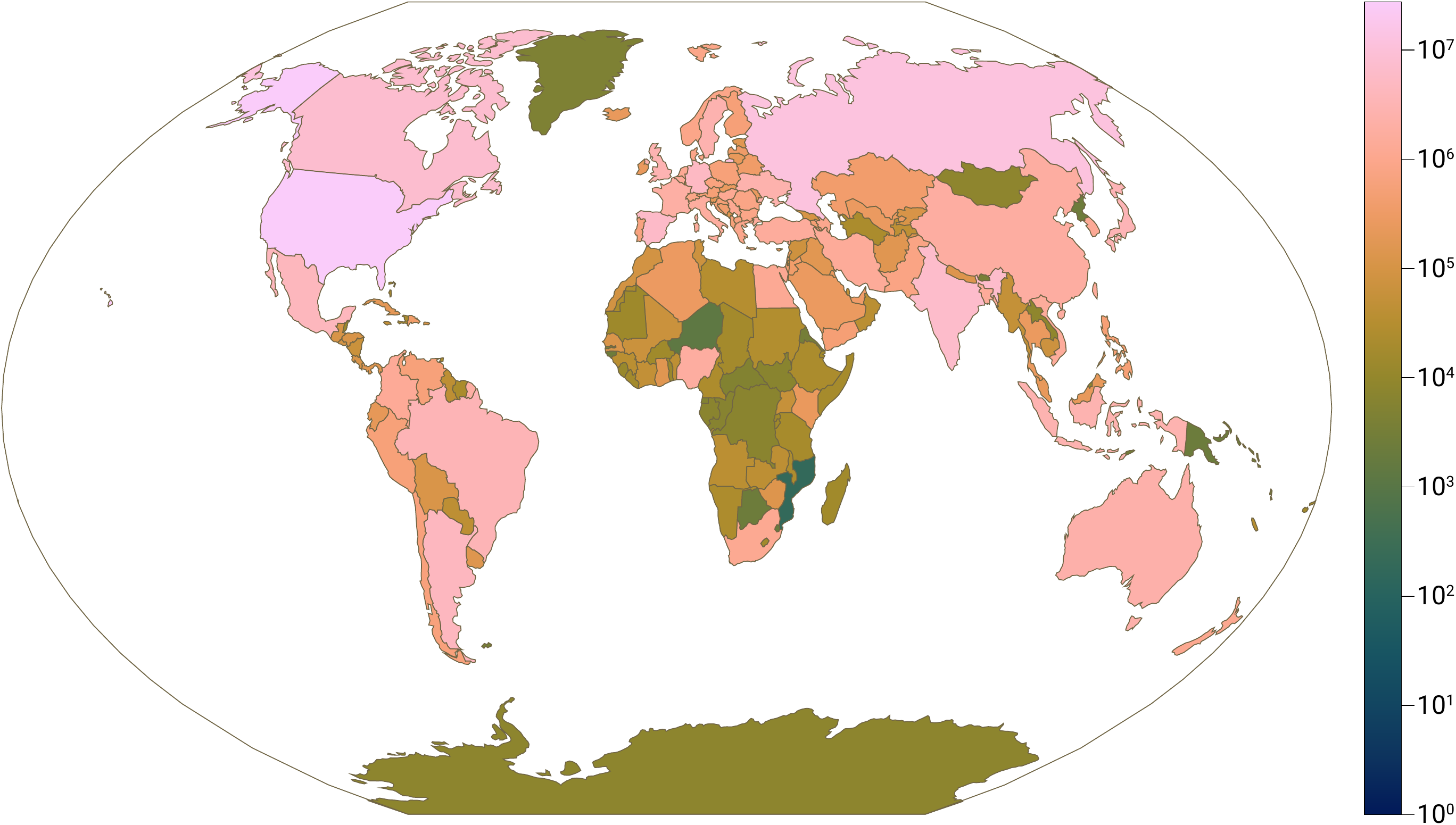}
    \caption{The number of stories for 2020 published from each country in Media Cloud, in log scale. The most stories ($10^{7.44}$) are from the U.S.}
    \label{fig:geo}
\end{figure}

\subsection{Geographic Media Collection}
Media Cloud exposes a breadth of geographic coverage that is unavailable in other tools, with extensive coverage outside of the U.S. (figure \ref{fig:geo}). The geographic media collections are a curated lists of the media sources published in a particular country. These were originally populated from a data ingest of the ABYZ News Links\footnote{\url{http://www.abyznewslinks.com/}} archive in 2018. Subsequently, our team has checked each collection for validity, and consulted with domain experts in many countries to improve coverage. The list of media sources from some countries have been less well vetted, but they are still present. A static snapshot of these geographic collections is available at the Harvard Dataverse at \url{https://doi.org/10.7910/DVN/4F7TZW}, which includes over 1.5 thousand geographic collections and 12.7 thousand associated sources. This dataset can be used to drive both research and other media analysis systems.  

New sources are added to the system through a variety of mechanisms. Because the media landscape is constantly changing, we continuously collaborate with local partners to validate and curate geographic collections. In addition, while Topic-specific data ingestion is underway for a research project, any new media source domains encountered will be added to the system as placeholders. Finally, system users outside of the core team can suggest media sources and collections to be added to the regular data ingest pipeline. These are vetted and approved by Media Cloud staff, based on overall system resources and validation of proposed sites as online news sources that write primarily about current events. Beyond collating by country of publication, we also specifically curate thematic collections related to research areas of interest: misinformation, reproductive health rights, partisan slant, etc.

\subsection{Voter Fraud Topic Data}
A key feature of Media Cloud Topic Mapper is its ability to generate large datasets of open web stories relevant to a particular topic.  For this paper, we include a sample dataset of stories, media, and links covering the vote-by mail-controversy in the U.S. during the 2020 presidential election. This dataset, available at the Harvard Dataverse at \url{https://doi.org/10.7910/DVN/D1XLWY},  was the basis for a paper that found that the primary drivers of voter fraud misinformation were President Trump and political and media elites rather than social media \cite{benkler2020}. 

The dataset was generated using the Topic Mapper with the Boolean query \texttt{(vote* or voti* or ballot) and (mail or absent*) and (fraud or rigg* or harvest*)}; the date range February 1, 2020 to November 23, 2020; and media sources from the following collections: United States - State \& Local, US Center Right 2019, US Left 2019, US Center 2019, US Right 2019, U.S. Top Sources 2018, US Center Left 2019, U.S. Top Digital Native Sources 2018, and U.S. Top Newspapers 2018. In addition, the dataset was seeded with URLs parsed from a 1.1 million tweets discovered by executing the same query over the same dates on Brandwatch. The resulting dataset includes 120,210 stories from 6,992 media sources, 82,636 story-to-story hyperlinks, and 33,029 media-to-media hyperlinks aggregated from the story links. 

\subsection{Generating Datasets}
Options for generating custom datasets are built into Media Cloud at all levels, including:
\begin{itemize}
\item Explorer: The Explorer tool allows the user to download all metadata for all stories matching any query executed on the system.
\item Topic Mapper: The Topic Mapper tool allows the user to download all metadata for all stories and media discovered through the topic query and spider process.
\item API: The open API allows programmatic download of metadata about all stories, media, feeds, topics, and tags in the system, including the ability to download word counts for any query, down to individual stories.
\item visualizations: almost every one of the hundreds of visualizations made available by the web tools provides a CSV download of the data underlying the visualization.
\end{itemize}

\subsection{Adherence to FAIR Principles}
The Media Cloud system adheres to the FAIR principles. As noted previously, we have uploaded our sample datasets to the Harvard Dataverse. In this paper, we describe the rich metadata and unique identifiers, while the upload snapshots serve as concrete examples. Datasets are accessible via the web using our published APIs and by downloading standard CSV files via the web apps. Easy access to story metadata is provided on our story details page.\footnote{\url{https://explorer.mediacloud.org/#/general/story/details}} In addition to the standard methods of consumption (e.g. API, CSV files), the data are made further interoperable via API clients available in Python\footnote{\url{https://github.com/mediacloud/api-client}} and \textsf{R}\footnote{\url{https://github.com/jandix/mediacloudr}}. To be \textit{reusable} by other researchers, our datasets and metadata must be clearly described, which we have done in this paper. A key tenet of \textit{reuse} is provenance, and in the case of stories this is represented by the included URLs.

\section{Key Areas of Use}
\begin{figure}[t]
    \centering
    \includegraphics[width=\columnwidth]{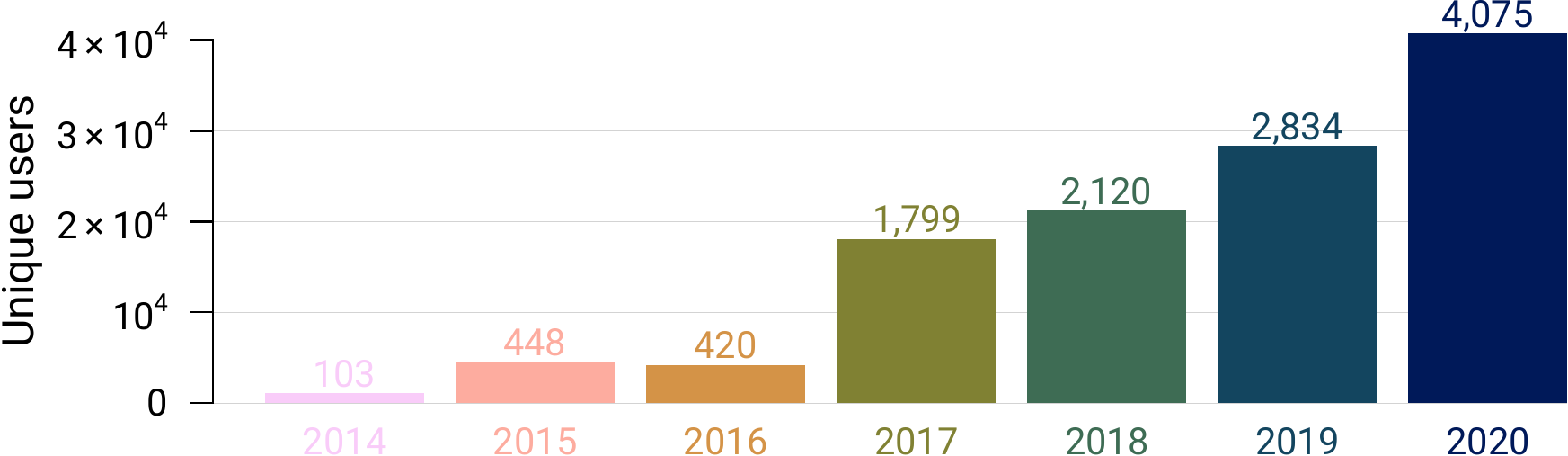}
    \caption{Yearly count of active (performing at least one action) unique users, excluding internal Media Cloud users.}
    \label{fig:users}
\end{figure}

\begin{figure}[t]
    \centering
    \includegraphics[width=\columnwidth]{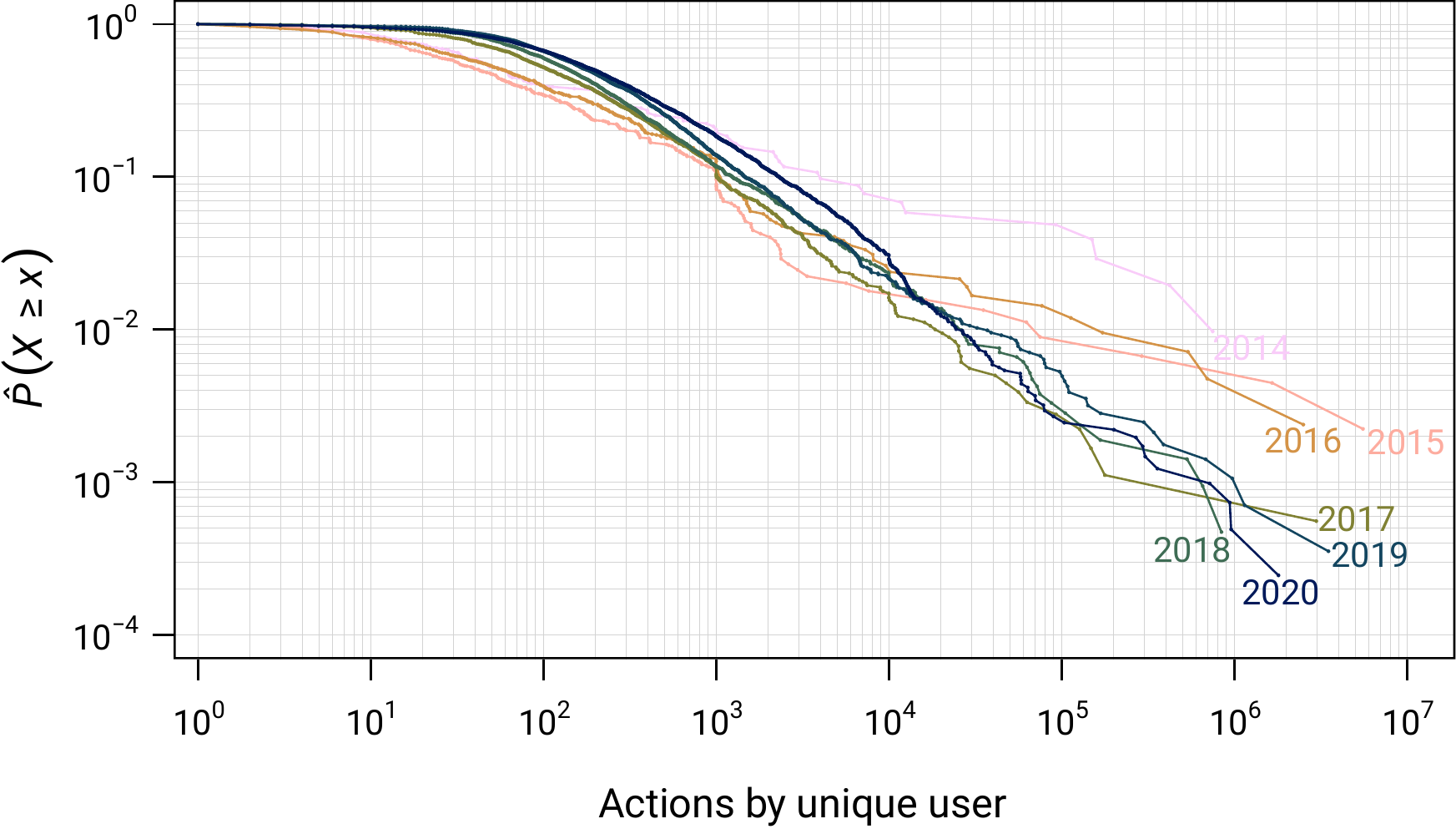}
    \caption{An empirical complementary cumulative distribution function (CCDF) for the count of actions by unique users, again excluding internal Media Cloud users, per year.}
    \label{fig:usage}
\end{figure}

Media Cloud fosters a community of researchers and technologists that engage with us via academic conferences, a dedicated Slack channel, a community email list, and directly via support emails. This community continues to grow year after year with over 4 thousand active (e.g., performed at least one action that year) users in 2020 (figure \ref{fig:users}). Some of these users, excluding internal Media Cloud users, performed hundreds of thousands of actions (figure \ref{fig:usage}). 

\begin{figure}[!ht]
    \centering
    \includegraphics[width=\columnwidth]{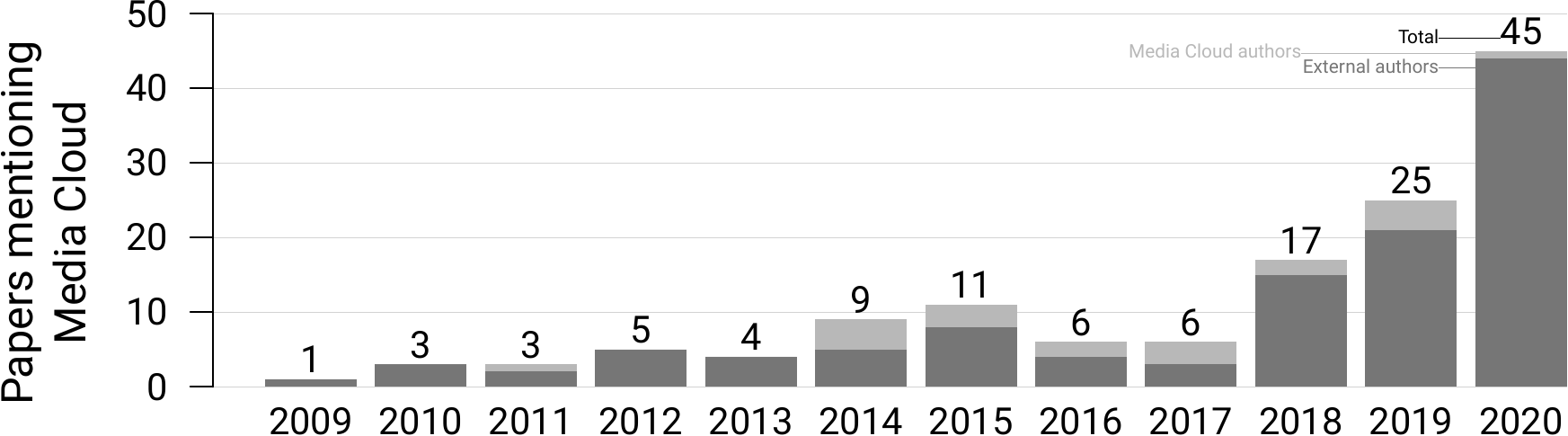}
    \caption{Count of mentions of mediacloud.org in scholarly articles published between 2009 to 2020, via a January 2021 search in Google Scholar. Articles by the Media Cloud team are shown separately in light gray.}
    \label{fig:mentions}
\end{figure}

As an academic engine for datasets used to support published research, Media Cloud has been growing in mentions since 2009. Figure (\ref{fig:mentions}) shows this pattern over time, with 45 papers mentioning Media Cloud in 2020 (only one of which was a publication from the Media Cloud team). 

Media Cloud supports researchers beyond the immediate Media Cloud team. As seen in figure (\ref{fig:mentions}), mentions of Media Cloud are on the rise, especially amongst external researchers. Recent papers mentioning Media Cloud include one describing it as a solution to archiving content from hyper-partisan sources \cite{tucker2018}, a recent study of COVID-19 news and its association with sedentary activity that uses Media Cloud \cite{huckins2020}, and a work suggesting that Media Cloud could be used to track the flow of ideas as applied to macroeconomics \cite{brazeal2011}.

Projects that use Media Cloud come from a diverse set of scholars, activists, journalists, and educators. Some areas of research Media Cloud datasets have already supported include:

\paragraph{Mapping national political discourse.} Media Cloud supports deep exploration of national media stratification, yielding significant findings of difference between countries. Researchers have documented asymmetric polarization in the U.S. media landscape \cite{benkler2018}, while in France researchers have found a split between institutionalists and the ``anti-elite'' \cite{zuckerman2019b}, but no evidence for a left/right split in influential news media.

\paragraph{Online mis/dis-information.} Media Cloud supports tracking narratives emerging from unreliable sources, surfacing coordinated attempts to mainstream untrue news stories, and investigating how media are manipulated to seed fear, uncertainty and doubt by those in political power \cite{benkler2020}. In the sub-domain of public health misinformation, our dataset has supported research into Ebola-related misinformation online \cite{roberts2017}, the spread of conspiracy theories related to water fluoridation in the United States \cite{helmi2018}, investigations into false claims about risk from casual exposure to fentanyl \cite{beletsky2020}, and mis/disinformation about climate activist Greta Thunberg \cite{dave2020}. 

\paragraph{Media self-analysis.} Media Cloud has been used as a tool for self-reflection within the journalism industry. Journalism schools have used the dataset to model the agenda of election coverage in the U.S. over the course of primaries for the U.S. presidential election \cite{bajak2019}, and journalists have similarly used Media Cloud to track and compare their own coverage of controversial stories \cite{rakich2020}. Our research team compared how different global outlets covered the 2019 Christchurch shooting against industry recommendations for covering terrorist attacks \cite{baumgartner2019}. A variety of teachers are using Media Cloud in higher-ed classroom settings to introduce approaches to studying media attention online using our highly accessible web tools.

\paragraph{Human rights and social justice media impact.} Many human rights foundations specifically fund non-profits to create impact on the media narratives around an issue they want to address. Media Cloud has helped these types of foundations and advocacy organizations track and measure media mentions \cite{meag2020}. Our own team has worked with Define American to analyze the language of news stories about immigration to America \cite{ndulue2019}, and analyzed media stories about police killings of unarmed people of color in the U.S., finding significant changes in the framing over time based on an analysis of news from before and after the killing of Michael Brown in 2014 \cite{zuckerman2019}. 

\paragraph{Media-based art and advocacy.} Artists have used the Media Cloud corpus to create interactive pieces showing news reporting on global conflicts \cite{bhargava2003}. Other researchers \cite{dignazio2020} are supporting advocacy groups across the Americas in their efforts to track gender-related killings of women by creating automated data pipelines built on Media Cloud to surface potentially related news stories.

\section{Discussion and Conclusion}
This paper introduces the Media Cloud platform, documenting how it is accessed, how it is produced, and how it is being used to generate open, accessible research datasets. Media Cloud provides global coverage, easy access, multiple levels of entry, and dataset creation for researchers studying attention, representation, influence and language in online news. Here, we discuss some logistical challenges, potential theoretical limitations, and future directions. 

First, content collection and curation on the open web grow more difficult \cite{bossetta2015}. As mentioned previously, RSS has long been a central element of the regular data ingest into Media Cloud, but it is less frequently supported. In response, we have taken efforts to increase the robustness of our content discovery via spidering and are also experimenting with ingesting sitemaps. In addition, tracking and maintaining the geographic media collections requires significant staff time and local domain expertise. 

Another barrier is the growth of news content surfacing on social media platforms. Studying the media ecosystem requires understanding how news content moves across these platforms alongside the social media users who are potential news consumers \cite{vandijck2013}. We are now beginning to directly index some platforms, including YouTube via their API, and incorporating independent research archives such as the Pushshift Reddit Dataset \cite{baumgartner2020}. This allows us to connect traditional stories with social sharing metrics of influence and also surface discrepancies between mainstream media and social sharing practices. 

We also wrestle with questions of whether the online media we index can be considered a reasonable proxy for the stories and topics a media source covers more broadly. Are the stories on the New York Times' website a full and accurate representation of the content that appears within the paper version of the newspaper? While print alone is becoming less important in overall news consumption \cite[The New York Times, for example, reported having 831 thousand print subscribers as compared to 4.7 million digital subscribers in 2020;][]{lee2020}, this challenge is still acute for television and radio news, which continue to be important sources of information for large parts of the population \cite{barthel2020}. For these, a given source's online content may not be a good proxy (e.g., how much talk radio is transcribed and available online?). We believe Media Cloud provides an accurate picture of the online presence of these sources, but acknowledge the lack of coverage of offline media as a potential gap in our methodology and it is something we seek to address.

A key issue with the meaningfulness of any collection is a \textit{denominator problem}: in order to know what percentage of online media is covering or discussing a topic one way or another, we need to know the total universe of online media. Practically, such exhaustive sampling is impossible both because some content is not openly available and because Media Cloud's ingest and spidering can ultimately only collect a tiny sample of the open web. Thus, instead of attempting to solve the denominator problem, Media Cloud focuses on \textit{prominence} using a relational approach, similar to IssueCrawler's claim that ``hyperlinking and discursive maps provide a semblance of given socio-epistemic networks on the web'' (\citeauthor{rogers2000} \citeyear{rogers2000}; see also \citeauthor{rogers2010} \citeyear{rogers2010}). Specifically, we assert that visibility via snowball link sampling, and then some form of centrality or threshold (e.g., above some minimum in-link degree, or in some upper percentile of in-link degree) within the resulting network, is a proxy for a story being prominent in the attention of media consumers. But the validity of such uses of the link economy to measure prominence relies on several key assumptions:
\begin{itemize}
    \item Stories hyperlink to other stories on which they rely or which they reference;
    \item Stories are influenced by the stories to which they link;
    \item The influence of stories summarizes the media landscape; 
    \item The media landscape determines the bounds of what people can consume; and
    \item The most prominent media items (the ``center'' of the media landscape) are the most influential on news consumers.
\end{itemize}

\noindent We have found the first assumption violated in many countries, notably in specific work on Saudi Arabian, Egyptian, Brazilian, Indian, and Colombian media, where web pages rarely link to external web pages. We have nonetheless had success analyzing influence in these non-hyperlinking counties by using alternative measures of influence such as social media metrics and third party panel-based metrics. 

For the last assumption, there is always the possibility of stories with a large influence on the public but not linked to by other media sources. This points to a larger, deeper issue: Media Cloud is only capable of measuring production, and not consumption. Our ultimate interest in media, especially as forming an essential pillar of democratic society, is: \textit{what media are having what influence, on whom?} But this is a question of consumption, not production. The connection between them is a long-standing concern in media research \cite[e.g.,][]{protess1992}, and divergences threaten the validity of any attempt to use media ecosystems to study civil society; e.g., \citeauthor{bruns2007} (\citeyear{bruns2007}) pointed out for IssueCrawler that links (production) do not say anything about the traffic that may go over those links (consumption). Even studying tweets is evidence only of what some Twitter users produce, providing only indirect evidence of influence on those users, who themselves are a tiny and biased sample of citizenry. 

It ultimately falls to users of Media Cloud to think through if and how its collection of online media provides a sufficiently valid way to measure a construct of interest, just as we are constantly doing internally. Our hope is that by being transparent about our methodology, it is easier for other researchers to reason through these issues as well. 

The technology maintenance itself also poses challenges for the project. Our corpus of stories continues to grow and stress the underlying storage architectures. Keeping a collection of almost 2 billion news stories highly available for instant search results, while evolving the research it supports, requires significant investments in computational hardware and personnel. Then, although it is a positive thing that Media Cloud is seeing increased usage by researchers of many backgrounds, this adds additional stress. Lastly, our project has been generously supported by foundations to research specific topics, but we have not raised long-term funding to maintain the platform as an open resource for the broader research community, even though the core team believes strongly in that goal. 

Despite these challenges, the growing user-base and mentions in scholarly literature demonstrate Media Cloud's widespread usefulness, and attest to its increasing visibility and recognition. We hope that this description will support Media Cloud's continued growth, empower more researchers to study the online media ecosystem, and inspire new and creative uses of Media Cloud. 

\section{Acknowledgments}
We thank Becky Bell and Jason Baumgartner. We also thank the collection curators. 

\fontsize{9.0pt}{10.0pt} \selectfont
\bibliography{mediacloud2021.bib}

\newpage
\section{Appendix}
In this online appendix, we include a description of data schema for tables in the Media Cloud database, and screenshots of the three web-based research tools. 

\subsection{Schema}
The Media Cloud data schema consists of roughly 150 tables in the database. Below, we describe the externally most relevant. 

\subsubsection{Content}
Media Cloud stores the content of each story as full HTML, as extracted text, and as individual sentences. To avoid violating publisher copyrights, Media Cloud does not share this content directly with the public. However, the system makes unordered word counts available for all queries on its system, down to the individual story. A user can request word counts for a media source for a year or more, for all stories mentioning a keyword, or for each story within any query returned by the system. Media Cloud also stores URLs for all stories indexed so that researchers can collect stories returned by any search via \texttt{wget} or a web browser. 

\subsubsection{Stories}
Each story within Media Cloud represents a single piece of content published on the web. Stories are associated with a single media source. Each story includes the following metadata: %usually a story on a news-oriented website
\begin{itemize}
    \item \texttt{stories\_id}: the unique internal id
    \item \texttt{media\_id}: the internal id of the parent media source
    \item \texttt{title}: the title, generated either through structured syndicated data or guessed from the HTML
    \item \texttt{publish\_date}: the publication date, generated either through structured syndicated data or inferred from the HTML
    \item \texttt{collect\_date}: the date the system collected the story
    \item \texttt{url}: the URL used to download the story
    \item \texttt{guid}: the global unique identifier for the story, generated either through structured syndicated data or guessed to be the same as the URL
    \item \texttt{language}: the language of the story text, generated by the cld2 library
    \item \texttt{tags}: tags associated with this story (see below)
\end{itemize}

\subsubsection{Media}
Each media source in Media Cloud represents a distinct publisher of data, such as The New York Times or Breitbart. Each media source includes the following metadata:
\begin{itemize}
    \item \texttt{media\_id}: unique internal id
    \item \texttt{name}: unique human readable name, generated by manual curation or guessed to be the URL or the HTML title of the media source home page
    \item \texttt{url}: URL of home page
    \item \texttt{start\_date}: the date the system starting collecting data for this media source
    \item \texttt{tags}: tags associated with this media source (see below)
\end{itemize}

\subsubsection{Feeds}
Each feed in Media Cloud represents an RSS, RDF, Atom, or other similar syndicated feed. Feeds are associated with a single media source. Each feed includes the following metadata:
\begin{itemize}
    \item \texttt{feeds\_id}: unique internal id
    \item \texttt{media\_id}: internal id of parent media source
    \item \texttt{url}: URL of feed resource
    \item \texttt{active}: Boolean indicating whether feed is still collected by the syndicated feed crawler
    \item \texttt{type}: `syndicated' or one of a number of internal types that indicate the feed is used as a virtual container for stories generated by something other than a syndicated feed, such as the topic spider
\end{itemize}

\subsubsection{Tags}
Media Cloud uses tags to associate various sets of metadata with stories and media. The same set of tags can apply to either stories or media. Each tag belongs to a tag set, which acts as a namespace for a given set of tags. Each tag set name is unique, and each tag is unique within a given tag set. Each tag set includes:% the following pieces of metadata:
\begin{itemize}
    \item \texttt{tag\_sets\_id}: unique internal id
    \item \texttt{name}: unique name
    \item \texttt{label}: unique label suitable for UI display
    \item \texttt{description}: short description of the tag set
\end{itemize}

\noindent Each tag includes the following metadata:
\begin{itemize}
    \item \texttt{tags\_id}: unique internal id
    \item \texttt{tag\_sets\_id}: internal id of parent tag set
    \item \texttt{tag}: unique name
    \item \texttt{label}: unique label suitable for UI display
    \item \texttt{description}: short description of the tag
\end{itemize}

\noindent Some of the most important tag sets (with [\texttt{tag\_sets\_id}]) are: 
\begin{itemize}
    \item \texttt{collection} [\texttt{5}]: curates media sources into sets useful for research, such as U.S. Top Online Media 2017, generated manually by the Media Cloud team
    \item \texttt{geographic\_collection} [\texttt{15765102}]: curates media sources into sets useful for international research, initially based on ABYZ News Links then manually curated% by the Media Cloud team
    \item \texttt{pub\_country} [\texttt{1935}]: marks primary publication location of a media source, generated manually by the Media Cloud team
    \item \texttt{nyt\_labels} [\texttt{1963}]: tags stories with thematic labels based on a machine learning model trained on the Google Word2Vec and NYT Annotated corpuses
    \item \texttt{primary\_language} [\texttt{1969}]: associates each media source with a primary language, generated automatically based on 50\% of stories detected to be in the primary language
    \item \texttt{twitter\_partisanship} [\texttt{15765109}]: groups media sources into U.S. political partisanship categories, based on the political affiliation of users who tweet each media source
    \item \texttt{cliff\_organizations} [\texttt{1988}]: assigns organization entities to stories, generated by the Cliff-Clavin system
    \item \texttt{cliff\_people} [\texttt{1989}]: assigns person entities to stories, generated by the Cliff-Clavin system
    \item \texttt{mc-geocoder} [\texttt{1989}]: assigns geographic areas of focus to stories, generated by the Cliff-Clavin system
\end{itemize}

\subsection{Screenshots}
\begin{figure*}[!ht]
    \centering
    \frame{\includegraphics[width=7in]{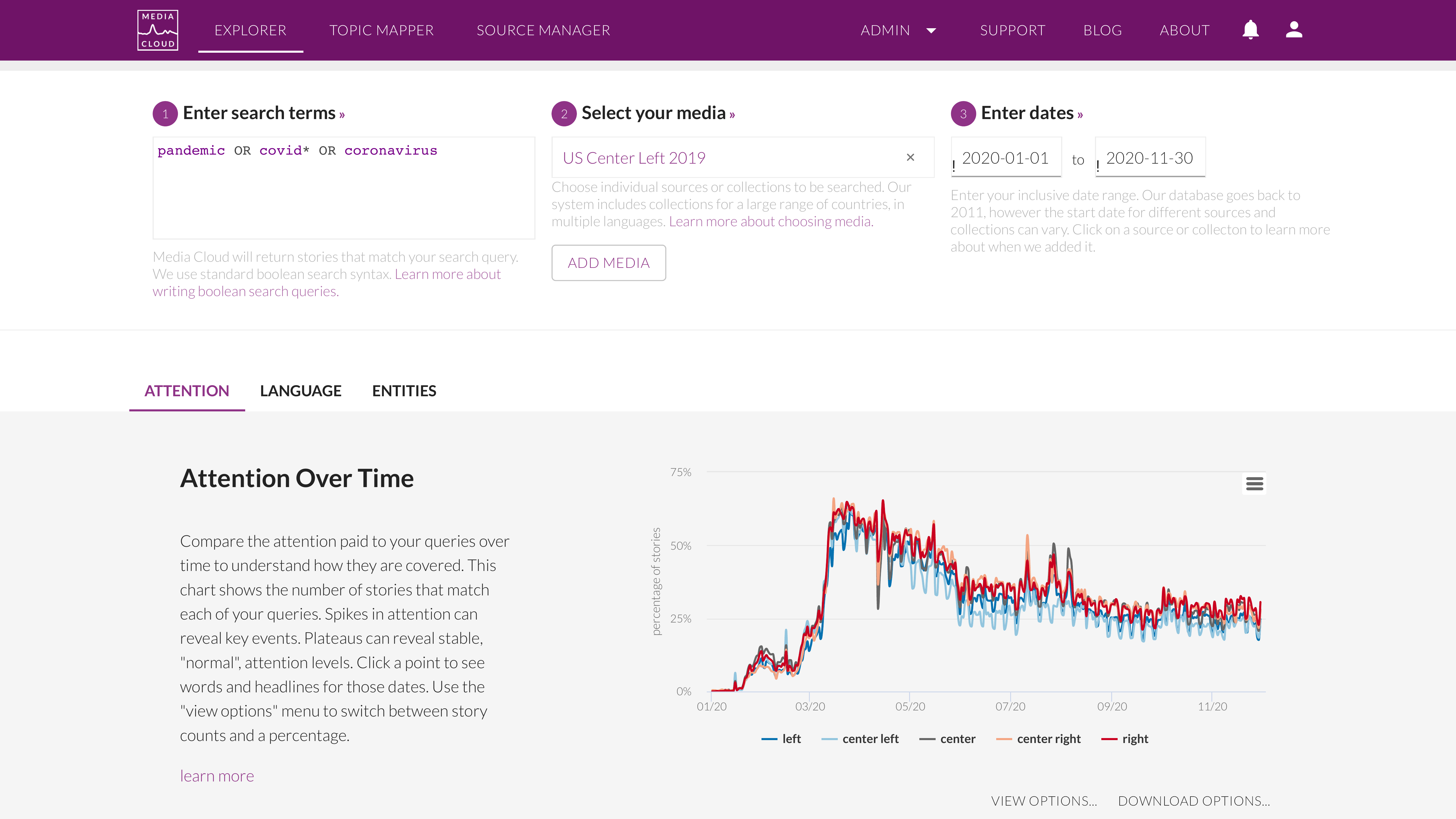}}
    \caption{A view of the Explorer tool (\url{https://explorer.mediacloud.org/}) showing a COVID-related search for 2020 across five pre-made collections of of U.S. media, showing the percentage of overall coverage (`attention') devoted to COVID-19 within each of those collections.}
    \label{fig:explorer}
\end{figure*}

\begin{figure*}[!ht]
    \centering
    \frame{\includegraphics[width=7in]{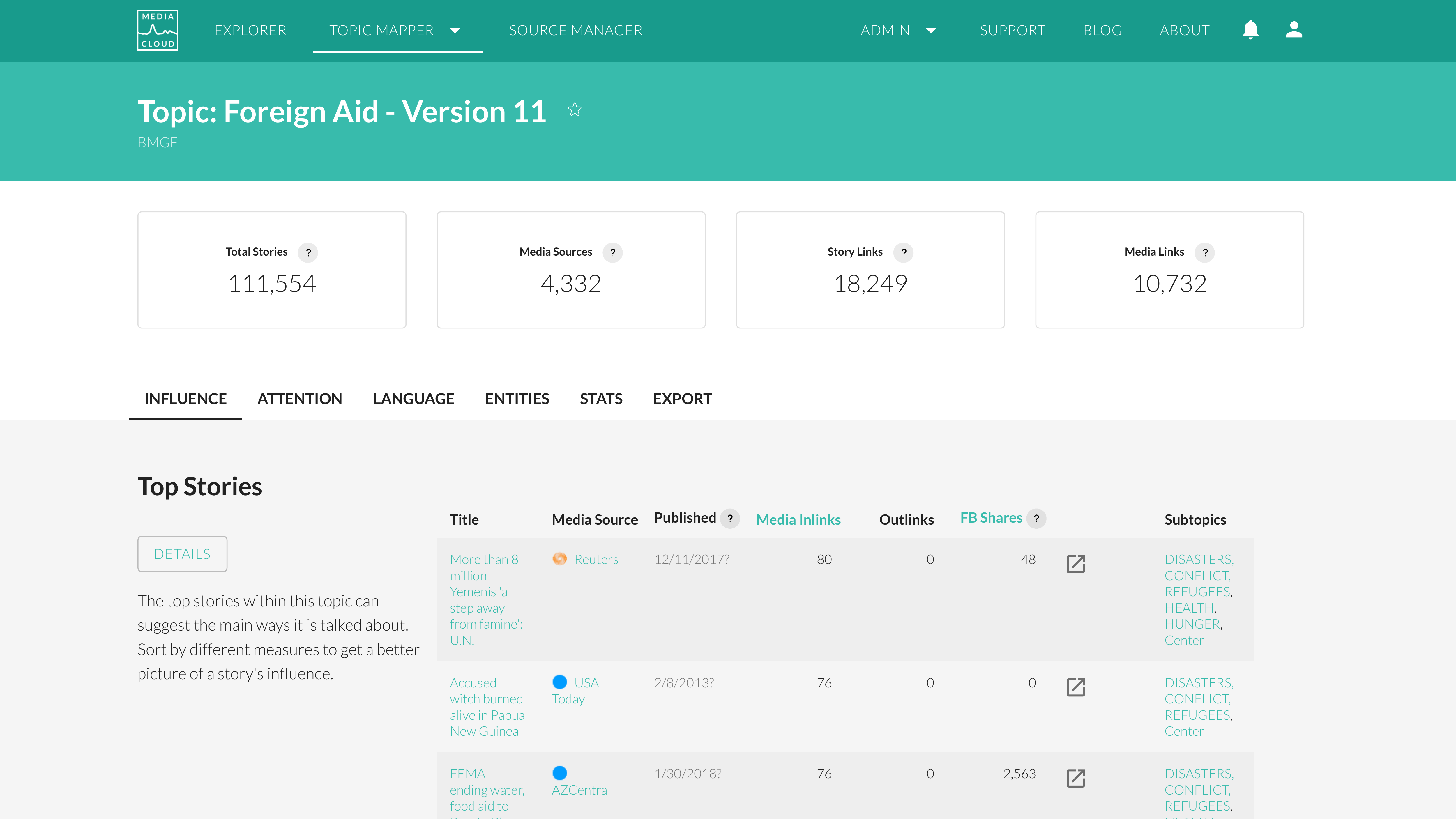}}
    \caption{A view, from December 21, 2020, of the Topic Mapper tool (\url{https://topics.mediacloud.org/}) showing the ``Foreign Aid'' topic.}
    \label{fig:topicmapper}
\end{figure*}

\begin{figure*}[!ht]
    \centering
    \frame{\includegraphics[width=7in]{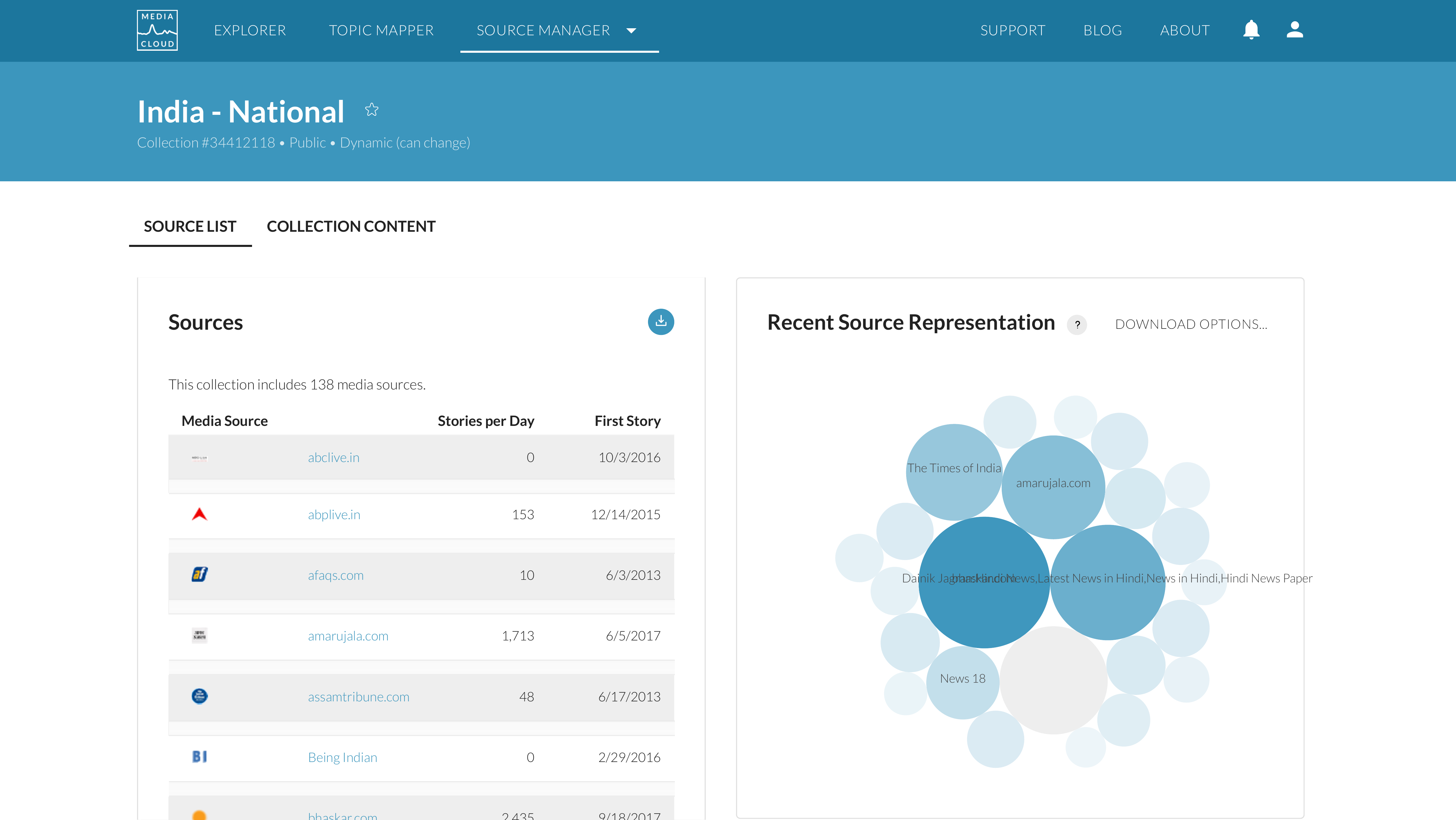}}
    \caption{A view, from December 27, 2020, of the Source Manager tool (\url{https://sources.mediacloud.org/}) showing sources for the ``India - National'' collection.}
    \label{fig:sourcemanager}
\end{figure*}

Below are screenshots taken in December 2020 of Media Cloud's web-based research tools: Explorer (\url{https://explorer.mediacloud.org/}) in figure (\ref{fig:explorer}), Topic Mapper (\url{https://topics.mediacloud.org/}) in figure (\ref{fig:topicmapper}), and Source Manager (\url{https://sources.mediacloud.org/}) in figure (\ref{fig:sourcemanager}).

\end{document}